# BlockAMC: Scalable In-Memory Analog Matrix Computing for Solving Linear Systems

Lunshuai Pan[1,2], Pushen Zuo[1,2], Yubiao Luo[1,2], Zhong Sun[1,2,3*], Ru Huang[1,2,3]

[1]School of Integrated Circuits  [2]Institute for Artificial Intelligence, Peking University

[3]Beijing Advanced Innovation Center for Integrated Circuits

* Corresponding author: zhong.sun@pku.edu.cn

*Abstract*—Recently, in-memory analog matrix computing (AMC) with nonvolatile resistive memory has been developed for solving matrix problems in one step, *e.g.*, matrix inversion of solving linear systems. However, the analog nature sets up a barrier to the scalability of AMC, due to the limits on the manufacturability and yield of resistive memory arrays, non-idealities of device and circuit, and cost of hardware implementations. Aiming to deliver a scalable AMC approach for solving linear systems, this work presents BlockAMC, which partitions a large original matrix into smaller ones on different memory arrays. A macro is designed to perform matrix inversion and matrix-vector multiplication with the block matrices, obtaining the partial solutions to recover the original solution. The size of block matrices can be exponentially reduced by performing multiple stages of divide-and-conquer, resulting in a two-stage solver design that enhances the scalability of this approach. BlockAMC is also advantageous in alleviating the accuracy issue of AMC, especially in the presence of device and circuit non-idealities, such as conductance variations and interconnect resistances. Compared to a single AMC circuit solving the same problem, BlockAMC improves the area and energy efficiency by 48.83% and 40%, respectively.

*Keywords—In-memory computing, analog matrix computing, RRAM, linear system, scalability.*

## I. INTRODUCTION

Solving linear systems has been playing an important role in modern scientific computing and other data-intensive tasks. In conventional digital computers, however, matrix computations are resource-demanding and time-consuming. Elaborated algorithms (direct or iterative) are used which generally are computationally expensive, featuring a polynomial time complexity, *e.g.*, $O(N^3)$, where $N$ is the size of matrix [1]. In the era of big data, the rapidly increasing data volume has imposed a strong challenge for the data processing of matrix in digital computers [2]. Additionally, since all digital computers adopt the von Neumann architecture, the data transfer bottleneck limits the computing throughput and energy efficiency of matrix computations [3].

Recently, in-memory AMC with nonvolatile resistive memory has gained extensive attention, because of its high speed and low complexity [4]. By configuring proper connections between the memory array and the peripheral amplifier circuits, it is possible to perform basic matrix operations in one step [5]-[8], such as matrix-vector multiplication (MVM) and matrix inversion (INV). This is enabled by the immense spatial parallelism in the crosspoint architecture and the concurrent feedback mechanism in the circuit. Theoretical analysis shows that the time complexity of in-memory AMC can be optimized to approach $O(1)$, which suggests that matrix problems with arbitrary sizes should be solved within a constant time [9].

In practice, the size of matrix mapping is limited by the manufacturability and yield of resistive memory array. The array size is generally below 256×256, in the consideration of multi-bit storage capability [10]. The memory cells may get stuck in the ON or OFF state, losing the tunability of conductance states [11]. On the other hand, as the matrix size increases, more hardware resources should be proportionally deployed in the periphery to perform AMC operations. Additionally, in a large-scale AMC circuit, the sources of non-ideal factors, such as memory cell variations and interconnect resistances, increase quadratically, thus causing significant degradation of computing accuracy [12]. As a result, it is challenging to implement large-scale AMC circuits. For MVM, it is convenient to divide a large-scale matrix into multiple sub-matrices, perform partial MVM operations with multiple memory arrays, and recover the final solution [13]-[15]. Such a divide-and-conquer method had also been implemented for INV computations in digital domain [16], by using multiple processors to enhance the computational parallelism. However, the fundamental limit of algorithmic complexity cannot be overcome, and the von Neumann bottleneck issue persists. In the case of INV in AMC, it remains unexplored.

In this work, we developed a method, termed BlockAMC, for solving large-scale linear systems with crosspoint resistive memory arrays. The key contributions of this work include:

- A compact algorithm for solving large-scale matrix inversion problem, by partitioning the original matrix as block matrices, and performing INV or MVM with each of them sequentially.
- The BlockAMC macro designs for operating with memory arrays, including an array of transmission gates for reconfiguring the circuit connections, a clock controller for scheduling it, a set of operational amplifiers for sharing to perform INV or MVM, sample-and-hold (S&H) circuits for storing and conveying intermediate results, as well as analog-digital conversion (ADC) and digital-analog conversion (DAC) interfaces. Both one-stage and two-stage BlockAMC have been implemented.
- The area and energy efficiencies of BlockAMC are analyzed, demonstrating substantial improvements over the original AMC circuits. The intrinsic conductance variations and interconnect resistances in the array are concerned, showing an improvement of computing accuracy, thanks to the reduced array size and error accumulation.

## II. ANALOG MATRIX COMPUTING CIRCUITS

There is a family of resistive memory concepts, such as resistive random-access memory (RRAM), phase-change memory (PCM), magnetoresistive RAM (MRAM), ferroelectric tunneling junction (FTJ) and ferroelectric field-effect transistor (FeFET), whose common features include the nonvolatile and reconfigurable conductance state of memory cell, and the crosspoint topology of the RAM architecture [17]. Nonetheless, RRAM stands out as a strong candidate, because of its prominent analog conductance characteristics, proper working voltage/current ranges in line with nowadays CMOS circuits and systems, in addition to the excellent scalability, fast write/read speed, and CMOS fabrication compatibility [13]. Note that although the industry-ready RRAM macros support only binary (single-bit) storage [19], many research works have demonstrated the analog conductance capability of RRAM devices, which is equipped with the verify scheme for precise tuning, and supported by the underlying physical mechanism that allows continuously modifying the internal state variable of the device [20]-[21]. In the following, we will assume analog RRAM for performing in-memory AMC, and the inherent conductance variations of RRAM devices will also be considered. First, we give a brief introduction to the two AMC primitives, namely MVM and INV.

*A. MVM Circuit*

Fig. 1(a) shows the MVM circuit, where a crosspoint RRAM array contributes a conductance matrix $G$ in real space. Voltages applied to the columns (bit-lines, BLs) of the array constitute a vector $v_{in}$. According to Ohm's law and Kirchhoff's current law (KCL), the currents collected on the WLs are the result of $G \cdot v_{in}$. Accordingly, the output voltages of TIAs dictate the resulting vector of MVM, namely $v_{out} = -\frac{G}{G_0} \cdot v_{in}$, which is a precise mapping of the following matrix operation:

$$b = Ax, \quad (1)$$

where $A$ is a matrix equal to $G/G_0$, $x$ is the input vector represented by $-v_{in}$, and $b$ is the output vector represented by $v_{out}$. Upon the application of $v_{in}$, the computation is triggered and the MVM result is obtained in one step. It is shown that computing time of the circuit is linearly dependent on the maximal sum of conductance along a row in the array, also controlled by the feedback conductance and gain-bandwidth product (GBWP) of TIAs [22].

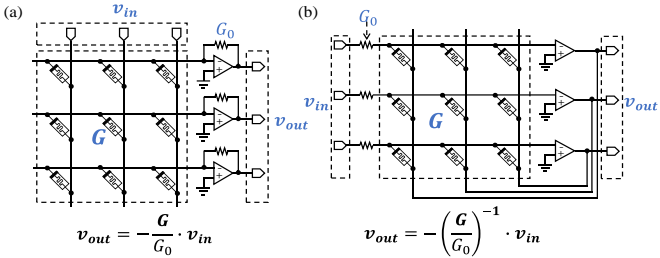

Fig. 1. In-memory AMC circuits for (a) MVM, and (b) INV operations

*B. INV Circuit*

Fig. 1(b) shows the INV circuit, which is used to solve a linear system:

$$Ax = b, \quad (2)$$

where $A$ is an invertible square matrix (assuming $n \times n$), $b$ is a known vector, and $x$ is the unknown vector to be solved. Mathematically, the solution is $x = A^{-1}b$, where $A^{-1}$ is the inverse matrix of $A$. Eq. (2) is the inverse problem of MVM in Eq. (1). It is interesting to note that the INV circuit is composed of exactly the same components as the MVM circuit, but configured with a different topology to perform the opposite function. The crosspoint RRAM array maps matrix $A$, but the resistors with conductance $G_0$ are used to convey the input voltage vector $v_{in}$, which maps the known vector $b$, contrary to the MVM case. The output voltages of operational amplifiers (OPAs) at equilibrium constitute the vector $v_{out}$, which is fed back to the BLs, forming nested feedback loops through the RRAM devices. According to Ohm's law and KCL, the currents collected on the virtual ground WLs are described by the equation $G_0 v_{in} + G \cdot v_{out} = 0$. To satisfy this equation, there must be $v_{out} = -(\frac{G}{G_0})^{-1} \times v_{in}$, which is the precise mapping of $x$. With this circuit, again, the matrix inversion problem is solved in one step. Its time complexity is also similar to the counterpart of MVM circuit, related to the minimal eigenvalue of an associated matrix and the GBWP of OPAs [23].

In Fig. 1, the RRAM cell is illustrated as a two-terminal, passive 1R structure for simplicity. In real world, usually the active one-transistor-one-resistance (1T1R) structure is adopted, to better control the programming of memory cells, and to eliminate the sneak current path issue [10], [11], [21], [24]. Additionally, the circuits in Fig. 1 are for AMC of positive matrix, as the device conductance can only be positive. For a matrix $A$ that contains both positive and negative elements, it should be split as $A = A_+ - A_-$, where $A_+$ and $A_-$ are both non-negative matrices, and are mapped in two memory arrays, respectively. In the hardware implementation, this is achieved by column-wise or row-wise splitting, together with the use of analog inverters or inherent differential input characteristics of OPA for the minus operation [4]. In both MVM and INV circuits, the result is accompanied with a minus sign, contributed by the negative feedback amplifiers.

Towards large-scale matrix computations, where the matrix cannot be accommodated in a single crosspoint RRAM array due to the manufacturability, yield, and non-ideality issues, matrix partitioning method should be used. Since MVM is a forward computing problem, it is easy to do this by using multiple arrays [13]-[15]. However, the situation is totally different for the inverse problem, INV. In the following, we introduce BlockAMC, which is used for solving large-scale linear systems by partitioning the original matrix into smaller block matrices. The MVM and INV circuits are used as two primitives for the block matrices, to construct the solution of the original linear system. Note that, there is a generalized block-matrix circuit for AMC, which, however, is about re-organizing the RRAM arrays in the form of block matrix for any given AMC circuit [25].

### III. DESIGN AND IMPLEMENTATION OF BLOCKAMC

In light of linear algebraic skills [26], the BlockAMC algorithm is designed as follows.

*A. BlockAMC Algorithm and System Architecture*

As shown in Fig. 2, the original matrix $A$ with a size of $n \times n$ is divided into four $n/2 \times n/2$ matrices: $A_1$, $A_2$, $A_3$, and $A_4$.

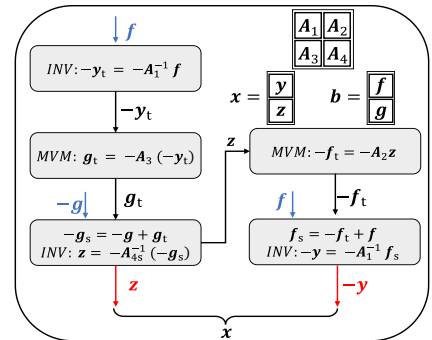

Fig. 2. BlockAMC flow chart, where the blue arrows represent the external inputs, the red arrows represent the output results.

Meanwhile, the input vector $b$ is divided into two $n/2 \times 1$ vectors: $f$ (upper) and $g$ (bottom). Here we assume $n$ is an even number, though the algorithm and implementation method are not limited only to such cases. For an odd $n$, size of $A_1$ can be chosen as $(n+1)/2 \times (n+1)/2$, and sizes of the other block matrices are consequently determined. Virtually, for a given matrix $A$, the size of $A_1$ can be arbitrarily selected, only requiring that it is square. To solve Eq. (2), the algorithm is carried out in five steps, consisting of three INV and two MVM operations, as summarized in the flow chart (Fig. 2).

1. INV operation with matrix $A_1$ and input vector $f$, obtaining the output vector $-y_t$, where the minus sign refers to the consideration of AMC circuit implementation.
2. MVM operation with matrix $A_3$ and input vector $-y_t$ cascaded from the first step, obtaining the output vector $g_t$, where the minus sign is removed.
3. INV operation with matrix $A_{4s}$ and input vector $-g_s$, both of which are defined as follows:

$$A_{4s} = A_4 - A_3 A_1^{-1} A_2, \quad (3)$$
$$-g_s = -g + g_t.$$

$A_{4s}$ is constructed with all the 4 block matrices. To be implemented in AMC, it should be calculated in advance, and stored in a crosspoint RRAM array, which may cause a pre-processing overhead. However, if either $A_2$ or $A_3$ is a zero matrix, the formula is reduced to $A_{4s} = A_4$. Consequently, it is the original $A_4$ that is stored in the memory array. This assumption holds in many practical cases, considering that the size of $A_2$ or $A_3$ may vary from 1 to $n$. The input vector contains the result $g_t$ cascaded from the second step, and the summation of $-g$ and $g_t$ can be conveniently achieved in the analog INV circuit. In this step, the output vector $z$ recovers the bottom part of the solution to vector $x$.

4. MVM operation with matrix $A_2$ and input vector $z$ cascaded from the third step, obtaining the output vector $-f_t$, showing again a minus sign.
5. INV operation with matrix $A_1$ and input vector $f_s$, the latter being $f_s = -f_t + f$. In this step, the concerned block matrix is same as the first step, resulting in a loop of cascading operations. The input vector is also similar, but containing the result of the last step. The output vector is $-y$, which recovers the upper part of the solution $x$.

**Algorithm 1** Divided Matrix computing for $Ax = b$

Input: $b = \begin{pmatrix} f \\ g \end{pmatrix}$ Divided Matrix: $A = \begin{pmatrix} A_1 & A_2 \\ A_3 & A_4 \end{pmatrix}$

Output: $x = \begin{pmatrix} y \\ z \end{pmatrix}$

1: Step1: $A_1 y_t = f$
2: function INV($A_1, f$)
3: $\quad -y_t = -A_1^{-1} f$
4: end function
5: Step2: $g_t = A_3 y_t$
6: function MVM($A_3, -y_t$)
7: $\quad g_t = -A_3(-y_t)$
8: end function
9: Step3: $A_{4s} z = g_s$, where $A_{4s} = A_4 - A_3 A_1^{-1} A_2$
10: function INV($A_{4s}, -g_s$), where $-g_s = -g + g_t$
11: $\quad z = -A_{4s}^{-1}(-g_s)$
12: end function
13: Step4: $f_t = A_2 z$
14: function MVM($A_2, z$)
15: $\quad -f_t = -A_2 z$
16: end function
17: Step5: $A_1 y = f_s$
18: function INV($A_1, f_s$), where $f_s = -f_t + f$
19: $\quad -y = -A_1^{-1} f_s$
20: end function

The workflow of the algorithm is illustrated in Algorithm 1. It consists entirely of linear matrix/vector operations, including MVM and INV, summation, and sign inversion, thus being highly compatible with the implementation of AMC circuits. The system architecture for implementing BlockAMC is shown in Fig. 3. For the four block matrices, MVM or INV operations are performed. Note that the $A_1$ array should be used twice, in the beginning and at the end of the algorithm. In this system, the operations and cascading are performed in a fully analog manner. To recover the solution of the linear system, the output results in steps 3 and 5 are collected and converted to the digital domain, through an ADC interface. Additionally, to receive the known vector $b$ as inputs in steps 1 and 3, a DAC interface is also designed to link the digital domain. The routes of input, output, and intermedia results are distinctly marked in Fig. 3.

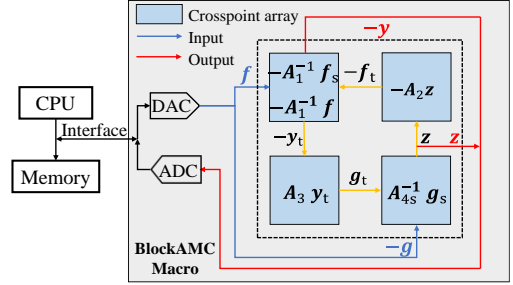

Fig. 3. System architecture, where the BlockAMC is connected to the data transmission bus. The blue, red, and yellow arrows indicate the input, output, and intermediate signals, respectively.

*B. BlockAMC Macro Design*

It has been shown that MVM and INV circuits are composed of the same components, including a crosspoint RRAM array and a column of OPAs, but only with different connections. Therefore, it is reasonable to propose a reconfigurable design to reuse the OPAs in a hardware macro, which is shown in Fig. 4. the macro design is composed of four crosspoint RRAM arrays storing the block matrices $A_1$, $A_2$, $A_3$, and $A_{4s}$, respectively. A set of OPAs for sharing to perform either MVM or INV operations, thus halving the count of OPAs compared to a single INV circuit. Compared to the single-function AMC circuits, the implementation of BlockAMC only incurs a small overhead of extra hardware deployments, including the necessary controller and reconfigurable connections between arrays.

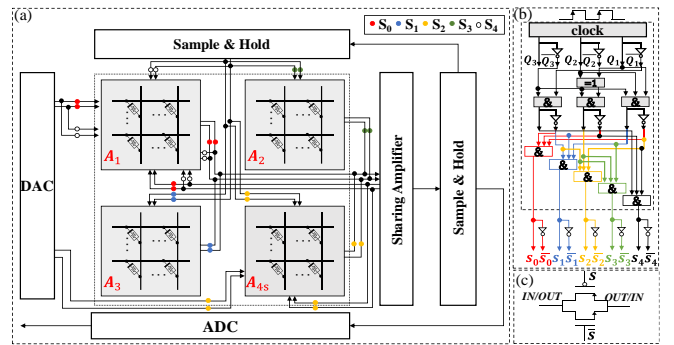

Fig. 4. BlockAMC macro design. (a) Detailed circuit design of BlockAMC, where the small circles with different color represent the on-off transmission gates and the black dots represent constant connections. (b) The functional controller design for the BlockAMC macro. (c) CMOS transmission gate.

By connecting the OPAs to a memory array successively, all MVM and INV operations can be performed to implement the BlockAMC algorithm. The connections between memory and OPAs are implemented by transmission gates, which are controlled by the clock signal. Fig. 4(b) shows the circuit design of the controller, which produces output control signals to turn on or off the transmission gates. Fig. 4(c) shows the CMOS transmission gates which are controlled by two signals in opposite phases $S$ and $\bar{S}$. In every clock cycle, an MVM or INV operation is accomplished. As shown in Fig. 4(a), there are five circuit connection topologies between a memory array and the OPAs, which are controlled by the transmission gates as marked by different colorful dots. Each set of transmission gates are turned on only in one clock cycle.

Between every two cascading AMC operations, the intermediate results are buffered by a set of S&H circuits and then conveyed to another set of S&H circuits to be provided as input. The use of two S&H modules renders the pipelining of the algorithm, thus improving the throughput of the system. In steps 1 and 3, the DACs, together with the S&H circuits, provide an input vector to the configured INV circuit. In steps 3 and 5, the output voltage of the INV circuit should be submitted to the ADC which converts the solution of the linear system to the digital domain.

### C. Two-stage solver

The implementation of BlockAMC can be progressively developed to solve ultra-large linear systems. Virtually, for an arbitrarily sized matrix, it can be partitioned stage by stage, resulting eventually in small scale block matrices that can be accommodated in memory arrays. Here we explain the case of a two-stage BlockAMC solver (Fig. 5). In the first step, the original large-scale matrix $A$ is divided into four $n/2 \times n/2$ first-stage block matrices as usual: $A_1$, $A_2$, $A_3$, and $A_{4s}$. According to the algorithm, INV operations should be performed with $A_1$ and $A_{4s}$. In the situation where $A_1$ and $A_{4s}$ are beyond the scale of AMC circuit implementation, they should be partitioned in the second stage, namely as $n/4 \times n/4$ block matrices. The first-stage matrices $A_2$ and $A_3$ will be used to perform MVM operations, they should also be partitioned to perform partial MVM and recover the results. In the two-stage solver architecture are included four one-stage BlockAMC macros, each performing INV or MVM and connected to the data bus to pipeline the algorithm. The output results in every one-stage BlockAMC macro are converted and stored in the main memory, which in turn will be converted back as analog input voltages for the following BlockAMC macro.

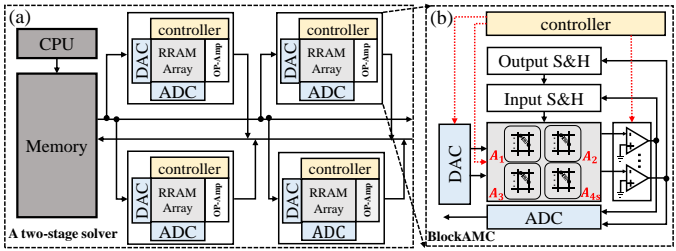

Fig. 5. Architecture of the two-stage BlockAMC solver.

## IV. SIMULATION RESULTS

The one-stage and two-stage BlockAMC macros have been designed at FreePDK 45 nm and validated in HSPICE to study the accuracy and performance. Each RRAM device is equivalent to a resistor with a specific conductance given by matrix mapping. Their performances are compared to the original single INV circuit that is composed of one large array, in the presence of two major device/circuit non-idealities, namely the device conductance variations and the interconnect resistances in the array. Without loss of generality, two popular types of matrices, namely the Wishart [27] and Toeplitz [28] matrices, are considered for the benchmarking, across a wide range of matrix size, from $8 \times 8$ to $512 \times 512$. It is recognized that AMC is hard to achieve high precision, rather it is positioned to provide a seed solution (or equivalently as a preconditioner) for digital computers, to speed up the convergence of iterative algorithms [29]. Consequently, an optimized AMC solution with higher computing accuracy should be beneficial to the acceleration of digital algorithms. Lastly, the circuit area and power consumption of different solvers are analyzed.

### A. Computing Accuracy Analysis

The Wishart matrix is widely referred to in fields such as statistical physics and engineering research. It is a stochastic matrix that can be generated by:

$$A = X^T X, \tag{4}$$

where vector $X$ is real Gaussian matrix with ($m \times n$) size, and $X^T$ is the transpose matrix of $X$. The size and element values of matrix $A$ are determined by the choice of $X$. Toeplitz matrix is a special cyclic matrix that is used in applications such as cyclic convolution and discrete Fourier [30]. Its form is as follows:

$$A = \begin{bmatrix} a_0 & a_{-1} & \cdots & a_{-n+1} \\ a_1 & \ddots & \ddots & \vdots \\ \vdots & \ddots & \ddots & a_{-1} \\ a_{n-1} & \cdots & a_1 & a_0 \end{bmatrix} \in R^{n \times n}, \tag{5}$$

where only elements in the first row/column are independent, and other elements along the diagonals are replicated.

Fig. 6 shows the results of an example in a numerical solver, HSPICE simulation results in the original AMC and BlockAMC, for the $256 \times 256$ Wishart matrix a random input vector, in the consideration of ideal mapping. During mapping, the matrix is normalized to make the largest element equal to 1. The resulting matrices are mapped to RRAM arrays, according to a unit conductance of $G_0=100$ μS. Fig. 6(a) shows the detail comparisons between the numerical results and the BlockAMC results from step 1 to step 5. Fig. 6(b) shows the final results of the numerical solver, original AMC, and BlockAMC. Fig. 6(c) shows the relationship between relative error and size of Wishart matrix, demonstrating a better computing accuracy of BlockAMC over original AMC in the ideal mapping case. The relative error is defined as

$$\varepsilon_r = \left| \frac{\sum_{i=1}^{n} \sqrt{(x_i - \hat{x}_i)^2}}{\sum_{i=1}^{n} \sqrt{x_i^2}} \right| \tag{6}$$

where $x_i$ and $\hat{x}_i$ represent the ideal and actual result of one element output, respectively.

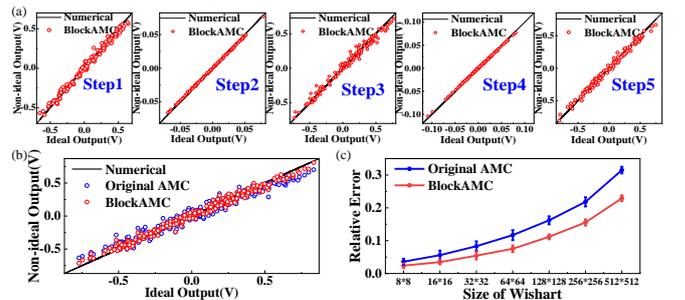

Fig. 6. (a) Comparisons between numerical and BlockAMC results in HSPICE step 1 to step 5. (b) The final comparison between numerical, original AMC, and BlockAMC for solving a $256 \times 256$ Wishart linear system, where the results of BlockAMC are from step 3 and step 5 in (a). (c) Computing accuracy comparison between original AMC and BlockAMC across a wide range matrix size.

Device variation is the most critical issue that impacts the accuracy of AMC. To compare the original AMC and BlockAMC, the device variation is assumed following Gaussian distribution, with a standard deviation of $0.05G_0$, which is achievable by using the write&verify algorithm [6], [20]. 40 random simulations were carried out for each matrix size. Fig. 7(a) and 7(b) show the results comparing the computational results of BlockAMC and original AMC for Wishart and Toeplitz matrices, respectively. In the case of Wishart matrix, the computing accuracy of two methods are almost identical across the range of matrix size, with the BlockAMC showing a slight improvement. In the case of Toeplitz matrix, the BlockAMC method shows a remarkable improvement, especially for the large-scale matrices. Therefore, in addition to solving the scalability issue of AMC, the BlockAMC may also provide a better seed solution with improved accuracy.

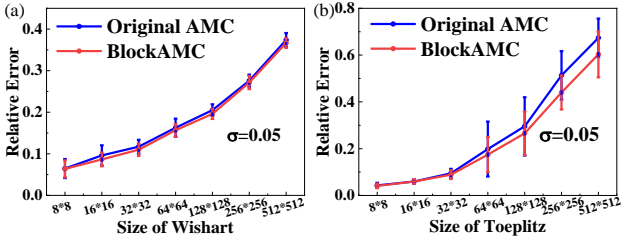

Fig. 7. Accuracy comparison between original AMC and BlockAMC with consideration of device variation, for (a) Wishart, and (b) Toeplitz matrix.

The two-stage BlockAMC solver has also been tested, for the $256 \times 256$ Wishart matrix for a parallel comparison. In this case, the original Wishart matrix is partitioned twice, resulting in 16 $64 \times 64$ block matrices for solving the linear system.

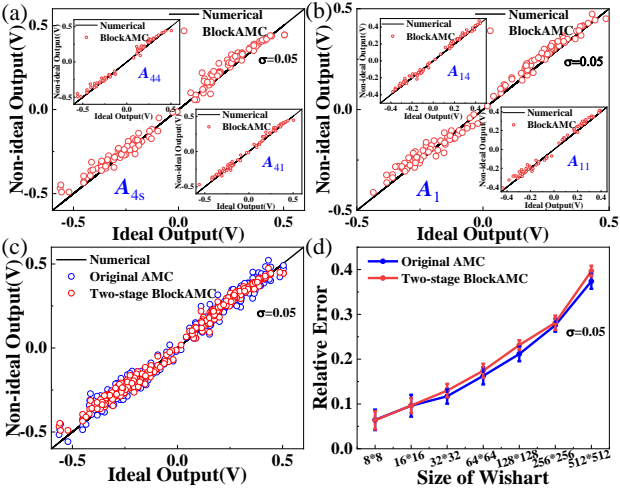

Fig. 8. Two-stage BlockAMC solver results for the $256 \times 256$ Wishart matrix with device variations. (a) Comparison between numerical and first-stage BlockAMC results of matrix $A_{4s}$ in HSPICE, which result from the second-stage BlockAMC solver of matrices $A_{41}$ and $A_{44}$. (b) Same as (a), but for block matrix $A_1$, which result from the second-stage BlockAMC solver of matrices $A_{11}$ and $A_{14}$. (c) The final comparison between numerical, original AMC, and BlockAMC. (d) Accuracy comparison between original AMC and two-stage BlockAMC for Wishart matrix of various sizes.

Fig. 8 shows the detailed simulation results in HSPICE. Fig. 8(a) and (b) shows the INV results of $A_{4s}$ and $A_1$ (step 3 and step 5, respectively) through the second-stage BlockAMC. In the insets, the INV results from their own block matrices are included. Based on these results, the final solution is summarized in Fig. 8(c). Fig. 8(d) summarizes the relative errors of original AMC and BlockAMC for Wishart matrix of various sizes, which is similar to the one-stage BlockAMC results in Fig. 7, supporting the scalability of this method towards larger scale INV problems through deeper partitioning.

Additionally, we have considered the effect of interconnect resistances in the comparison between the original AMC and one-stage and two-stage BlcokAMC. The segment resistance between every two memory cells along the BL or WL is assumed as 1 Ω, which is approximately the result in the 65 nm node [12]. The results in Fig. 9 show the one-stage BlockAMC has an advantage regarding computing accuracy over the original AMC approach, reducing the relative error by up to 10%. Compared to the one-stage solver, the two-stage BlockAMC enhances this improvement, thus demonstrating the superiority of this approach for scalable and reliable AMC implementations.

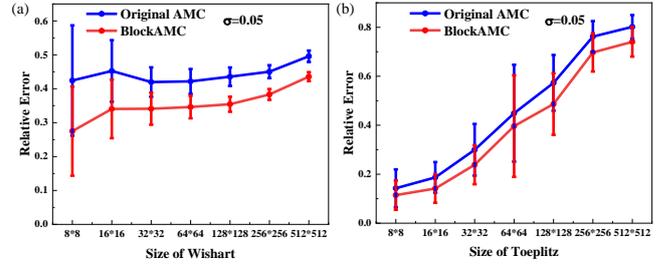

Fig. 9. Accuracy comparison between original AMC and BlockAMC with considerations of both device variation and interconnected resistance, for (a) Wishart, and (b) Toeplitz matrix.

### B. Macro Performance Analysis

The obvious advantage of BlockAMC is the reduced amount of hardware resources by reusing the peripheral circuits, such as OPA, DAC and ADC. Fig. 10 presents a quantitative estimation of area and power consumption of the three solvers, namely the original AMC, one-stage and two-stage BlockAMC solvers, for solving the same linear system with size of $512\times512$. It is composed of four major parts, contributed by OPA, DAC, ADC, and RRAM arrays, respectively. The contributions of other components in the macro, such as transmission gates and wires, are ignored. Specifically, the parameters for estimating the area and power of ADCs and DACs refer to previous works [30]. Those for OPAs and RRAM arrays are calculated with the parameters obtained in HSPICE. Particularly, the power consumption of OPAs is estimated by

$$P_{OPA} = NV_sI_q \qquad (7)$$

where $N$ is the number of OPAs, $V_s$ is the supply voltage, and $I_q$ is the quiescent current of OPAs [31].

The total areas of the three solvers are 0.01577, 0.00807, and $0.01383 mm^2$, respectively. Therefore, the proposed one-stage and two-stage BlockAMC solver saves 48.3% and 12.3% of chip area, respectively, than the original AMC. On the other hand, the one-stage and two-stage BlockAMC solvers and two-stage solver reduce 40% and 37.4% of power consumption than the original AMC solver, respectively. In the two-stage solver, OPAs are separately deployed for the first-stage INV and MVM, resulting in the same count of OPAs and thus a rise of area and power. Therefore, the BlockAMC method is effective to deliver a compact, energy-efficient design for solving large-scale linear systems.

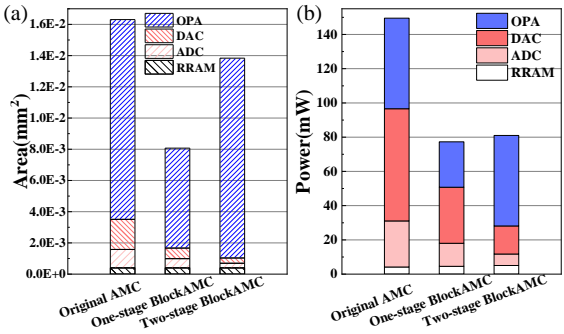

Fig. 10. The breakdown of (a) circuit area and (b) power consumption in three types of solvers.

## V. Conclusion

In conclusion, we have developed an effective method for AMC to solve large-scale linear systems by using multiple memory arrays whose size is available at the state of the art. According to the proposed algorithm, a one-stage and two-stage BlockAMC macros have been designed and validated, with considerations of device and circuit non-idealities. There exists benefits in terms of computing accuracy, chip area and energy improvements in BlockAMC than the original AMC with a single INV operation, thanks to the design of reuse of hardware resources as well as the pipelining dataflow. Hardware experiments with small-scale RRAM arrays are undergoing, as proof-of-concept demonstration of the BlockAMC method. We expect such a method should make a key contribution towards resolving the infamous scalability issue of analog computing in the modern times.

## Acknowledgment

This work was supported by National Key R&D Program of China (No. 2020YFB2206001), National Natural Science Foundation of China (Nos. 62004002, 92064004, 61927901), and the 111 project (No. B18001).